\documentclass[aps,preprint,showpacs]{revtex4}
\usepackage{times}
\begin{document}

\title{Circular Orbits under Central Forces in Special Relativity}

\author{
J M Aguirregabiria, 
A Hern\'{a}ndez 
and 
M Rivas 
} 
\affiliation{Dept.\ of Theoretical Physics and History of Science, 
The University of the Basque Country, \\
P.~O.~Box 644,
48080 Bilbao, Spain}

\begin{abstract} 
We discuss the existence and stability of circular orbits of a relativistic
point particle moving in a central force field. The stability condition is somewhat more
restrictive in Special Relativity. 
In the particular case of attractive central force fields
proportional  to $1/r^n$, 
we find the values the angular momentum may have in circular orbits
and show that, unlike in the non-relativistic limit, the latter are not always
stable  in Special Relativity for $2<n<3$.
\end{abstract} 


\maketitle

\section{Introduction}

Recently Torkelsson~\cite{tor} and Boyer~\cite{boy} have considered the
orbits  in a Newtonian force field $\mathbf{F}=-k\mathbf{r}/r^3$ in the
framework of Special Relativity. One of their results states that
circular orbits only exist for $L>k/c$, while in Galilean mechanics they 
exist for all non-zero angular momenta. At first sight one could
think that this is a restriction  on the possibility to have
relativistic circular orbits. The first goal of this note is to stress
that it is not so, for there is exactly one circular orbit for each
radius $r$ at which the force is attractive. In fact, the condition $L>k/c$
arises as a consequence of the interplay between the equation of motion
and the definition of the angular momentum in Special Relativity. We
will also show that similar restrictions on the values of $L$ (but not
of $r$) have to be satisfied by circular orbits in other central force
fields. Finally we will discuss how the familiar result~\cite{bb:mar} on the stability of
circular orbits in non-relativistic force fields  is changed by Special Relativity.
We check the stability condition with forces of the form
$\mathbf{F}=-k\mathbf{r}/r^{n+1}$ for constant $k>0$ and $n$, as well as with the Yukawa potential.

\section{Circular orbits and angular momentum}

Let us consider a central force field written as
 \begin{equation}
 \mathbf{F}=F(r)\,\frac{\mathbf r}r.
 \end{equation}
If a particle describes a circular orbit $r=r_0$ its equation of motion is
 \begin{equation}
 F\left(r_0\right)=-m\gamma\frac{v^2}{r_0},\qquad\gamma\equiv\left(1-\frac{v^2}{c^2}\right)^{-1/2}.
 \end{equation}
By defining the dimensionless quantity 
 \begin{equation}
 \mu\equiv-\frac{r_0F\left(r_0\right)}{2mc^2}>0,
 \end{equation}
one can check that the particle velocity and angular momentum are, respectively,
 \begin{eqnarray}
 v&=&c\,\sqrt{2\sqrt{\mu^4+\mu^2}-2\mu^2},\label{eq:v}\\
 L&=&mcr_0\,\sqrt{2\sqrt{\mu^4+\mu^2}+2\mu^2}.
 \end{eqnarray}
Since one can easily check from (\ref{eq:v}) that $0<v<c$ (and that $v$ increases 
monotonically with $\mu$), we conclude that \emph{there exists exactly one 
circular orbit for each radius $r_0$ at which the force is attractive}, $F\left(r_0\right)<0$: 
one simply has to select the corresponding velocity (\ref{eq:v})
in a direction perpendicular to the radius. But the fact that every radius is allowed 
does not necessarily mean that every angular momentum is allowed.

\subsection{Forces proportional to $1/r^n$}
In fact, if we choose, for constant $k>0$ and real $n$ a central force in the form
 \begin{equation}\label{eq:krn}
 F(r)=-\frac k{r^n}
 \end{equation}
one has for circular orbits
 \begin{equation}
 L=\frac k{\sqrt2\,c}\,\sqrt{r_0^{4-2n}+\sqrt{r_0^{8-4n}+\frac{4m^2c^4}{k^2}\,r_0^{6-2n}}}
 \end{equation}
and the following cases:
\begin{description}

\item[Case $-\infty<n<2$.] The angular momentum
 increases monotonically with the radius $r_0$ and all values $0<L<\infty$ are allowed.

In the particular case $n=1$, the velocity is independent of the radius, exactly as happens 
in the non-relativistic limit:
 \begin{equation}
 v=\sqrt{\frac{\sqrt{k^4+4k^2m^2c^4}-k^2}{2m^2c^2}},\quad \forall r_0.
 \end{equation}

\item[Case $n=2$.] For Newtonian forces one recovers the result of Torkelsson~\cite{tor}
and Boyer~\cite{boy}. There is a lower limit for the angular momentum (but not for the radius 
or the velocity):
 \begin{equation}
 L=\frac k{\sqrt2\,c}\,\sqrt{1+\sqrt{1+\frac{4m^2c^4}{k^2}\,r_0^2}}>\frac kc.
 \end{equation}
In the non-relativistic limit $c\to\infty$ the lower bound vanishes.

\item[Case $2<n<3$.] Since the angular momentum goes to $+\infty$ both for $r_0\to0$ and for 
$r_0\to\infty$, there must exist a minimum (positive) value. In fact, the minimum happens for
\begin{equation}
r_0=\left[\frac{\sqrt{n-2}\,k}{(3-n)mc^2}\right]^{\frac1{n-1}}
\end{equation}
and has the value 
 \begin{equation}\label{eq:lmin}
L_\mathrm{min}=mc\sqrt{\frac{3-n}{n-2}}\left[\frac{\sqrt{n-2}\,k}{(3-n)mc^2}\right]^{\frac1{n-1}}. 
 \end{equation}
In the non-relativistic limit $c\to\infty$ the lower bound vanishes
and in the limits $n\to2,\ 3$ we recover the neighboring cases.

\item[Case $n=3$.] There is also a lower limit for the angular momentum:
 \begin{equation}
 L=\frac k{\sqrt2\,c}\,\sqrt{\frac1{r_0^2}+\sqrt{\frac1{r_0^4}+\frac{4m^2c^4}{k^2}}}>\sqrt{km}.
 \end{equation}
In the non-relativistic limit, the angular momentum of all circular orbits is the same: $L=\sqrt{km}$.
In consequence, far more angular momentum values are allowed in the relativistic case.

\item[Case $n>3$.] The angular momentum
is monotonically decreasing with $r_0$. In consequence all positive values of $L$ are allowed.

\end{description}

\section{Stability of circular orbits}

If one uses the angular momentum conservation to eliminate from the equations of
motion the angular part, one obtains the equivalent one-dimensional problem.
Alternatively one can use the angular momentum 
\begin{equation}
L=m\gamma r^2\dot\varphi,\qquad\gamma\equiv\left(1-\frac{\dot r^2+r^2\dot\varphi^2}{c^2}\right)^{-1/2}
\end{equation}
to eliminate $\dot\varphi$ from the total energy
\begin{equation}
E=m\gamma c^2+V(r),\quad V(r)\equiv-\int F(r)\,dr
\end{equation}
and take the time derivative of the latter. In any case, the equation of the radial motion is
 \begin{equation}\label{eq:oned}
 \ddot r=f\left(r,\dot r\right)\equiv\frac{\alpha F(r)}{m\gamma_r^3}+c^2\frac{1-\alpha^2}{\gamma_r^2r},
 \end{equation}
where we have defined
\begin{eqnarray}
\gamma_r&\equiv&\left(1-\frac{\dot r^2}{c^2}\right)^{-1/2},\\
\alpha&\equiv&\left(1+\frac{L^2}{m^2c^2r^2}\right)^{-1/2},\qquad(0<\alpha<1).
\end{eqnarray}
Unlike in the Galilean case, the right hand side of (\ref{eq:oned}) depends on the velocity $\dot r$, so that
strictly speaking we do not have an effective potential (although a suitably generalization can be 
defined~\cite{tor}). However, this will not be an obstacle in the following.

Circular orbits are given by the conditions $r=r_0$ and $\dot r=0$  and satisfy
 \begin{eqnarray}
\gamma_r&=&1,\\
\alpha&=&\alpha_0
        =\sqrt{1+\mu^2}-\mu,\\
F\left(r_0\right)&=&-mc^2\frac{ 1-\alpha_0^2}{\alpha_0r_0}<0.
 \end{eqnarray}

To study the stability of the circular orbits, we will
consider a very near orbit in the form $r=r_0+\epsilon$, with $|\epsilon|\ll|r_0|$, so that
 in the linear approximation the equation of motion reads
  \begin{equation}\label{eq:linear}
\ddot\epsilon=\frac{\partial f}{\partial r}\left(r_0,0\right)\,\epsilon+\frac{\partial f}{\partial\dot r}\left(r_0,0\right)\,\dot\epsilon.
  \end{equation}
Since $\dot r$ only appears in $\gamma_r$ through its square, the last derivative in (\ref{eq:linear}),
which is proportional to $\dot r$, vanishes
and the equation of motion reduces to
 \begin{equation}
  \ddot\epsilon+\Lambda\epsilon=0,\quad\Lambda\equiv-\frac{\partial f}{\partial r}\left(r_0,0\right).
 \end{equation}
We are now in position to discuss the stability: if $\Lambda>0$ the orbit will oscillate around the circular one, but
if $\Lambda<0$ it will escape from the circular orbit, which will be unstable.
A straightforward calculation leads to
 \begin{equation}
 \Lambda=-\frac{\alpha_0}{mr_0}\,\left[r_0F'\left(r_0\right)+\left(2+\alpha_0^2\right)\,F\left(r_0\right)\right],
 \end{equation}
so that the stability condition reads
 \begin{equation}\label{eq:stc}
\frac{F'\left(r_0\right)}{F\left(r_0\right)}+\frac{2+\alpha_0^2}{r_0}>0. 
 \end{equation}
In the non-relativistic limit $c\to\infty$ one has $\alpha_0=1$ and we recover the familiar condition~\cite{bb:mar}
 \begin{equation}\label{eq:stc0}
\frac{F'\left(r_0\right)}{F\left(r_0\right)}+\frac{3}{r_0}>0. 
 \end{equation}

\subsection{Forces proportional to $1/r^n$}
In the family of force fields (\ref{eq:krn}) we get
 \begin{equation}
\Lambda=k\alpha_0\,\frac{1-\alpha_0^2}{mr_0^{n+1}}\,\left[(2-n)+\frac{m^2c^2r_0^2}{L^2}\,(3-n) \right],
 \end{equation}
so that we have the following cases:
\begin{description}

\item[Case $n\le2$.] Circular orbits are stable: $\Lambda>0$.

\item[Case $2<n<3$.] Circular orbits are stable for $r>\tilde r$ and unstable for $r<\tilde r$, with
 \begin{equation}\label{eq:rt}
  \tilde r\equiv\frac L{mc}\sqrt{\frac{n-2}{3-n}}.
 \end{equation}

\item[Case $n\ge3$.] Circular orbits are unstable: $\Lambda<0$.

\end{description}

In the limit $c\to\infty$ we recover the well known result~\cite{bb:mar}: 
circular orbits are stable for $n<3$ and unstable for $n\ge3$.

\section{Conclusions and comments}

We have shown that in a central force field there exists exactly one circular relativistic orbit for each radius
at which the force is attractive; but for $F(r)=-k/r^n$ with $k>0$ and $2 \le n < 3$ the angular momentum has 
the lower bound given by (\ref{eq:lmin}). For $n=3$ all $L>\sqrt{km}$ are allowed, while in the
non-relativistic case we always have $L=\sqrt{km}$.

As for the stability of circular orbits, the results in the relativistic and non-relativistic
cases are the same for $n\le2$ (stability) and for $n\ge3$ (instability); but for $2<n<3$ relativity
destroys stability  for any radius smaller than (\ref{eq:rt}). It is worth mentioning
that the Newtonian potential has the greatest $n$ for which all circular orbits are stable.

The last result suggests that relativistic corrections make more difficult the
stability of circular orbits. 
In fact, if one considers the Yukawa (or Coulomb shielded) potential 
\begin{equation}
V(r)=-\frac kr\,e^{-r/a},
\end{equation}
the stability condition (\ref{eq:stc}) reduces to
 \begin{equation}
 a^2+ar-r^2>\frac{k}{mc^2a^{1/2}}(a+r)^{3/2}\,e^{-r/a},
 \end{equation}
which is more restrictive that the non-relativistic condition~\cite{bb:mar}:
 \begin{equation}
 a^2+ar-r^2>0.
 \end{equation}
Something similar may happen in other cases, because if the relativistic condition (\ref{eq:stc})
is satisfied the non-relativistic (\ref{eq:stc0}) is also fulfilled, but the converse
is not necessarily true due to the fact that $\alpha_0<1$.

\acknowledgments
We gratefully acknowledge useful correspondence with Prof.\ T.\ Boyer.
This work was supported by The University of the Basque Country
(Research Grant~9/UPV00172.310-14456/2002).



\end{document}